\documentclass[a4paper,fleqn,usenatbib]{mnras}

\usepackage[T1]{fontenc}
\usepackage{ae,aecompl}

\usepackage{graphicx}	
\usepackage{amsmath}	
\usepackage{amssymb}	

\usepackage{hyperref}
\newcounter{ctr}
\def\ion#1#2{\setcounter{ctr}{#2}#1$\;${\small\Roman{ctr}}\relax}

\newcommand{\hal}{H\ensuremath{\alpha}}
\newcommand{\hbeta}{H\ensuremath{\beta}}

\newcommand{\fvar}{\ensuremath{f_\mathrm{var}}}

\title[No Narrow-Line Variability in Mrk 142]{No Evidence for
  [O~{\LARGE III}] Variability in Mrk 142}

\author[Barth \& Bentz]{
Aaron J. Barth,$^{1}$\thanks{E-mail: barth@uci.edu}
Misty C. Bentz$^{2}$\thanks{E-mail: bentz@astro.gsu.edu}
\\
$^{1}$Department of Physics and Astronomy, University of California,
Irvine, 4129 Frederick Reines Hall,
Irvine, CA 92697, USA \\
$^{2}$Department of Physics and Astronomy, Georgia State University,
Atlanta, GA 30303, USA
}

\pubyear{2016}

\begin{document}
\label{firstpage}
\pagerange{\pageref{firstpage}--\pageref{lastpage}}
\maketitle

\begin{abstract}
Using archival data from the 2008 Lick AGN Monitoring Project, Zhang
\& Feng (2016) claimed to find evidence for flux variations in the
narrow [\ion{O}{3}] emission of the Seyfert 1 galaxy Mrk 142 over a
two-month time span. If correct, this would imply a surprisingly
compact size for the narrow-line region. We show that the claimed
[\ion{O}{3}] variations are merely the result of random errors in the
overall flux calibration of the spectra.  The data do not provide any
support for the hypothesis that the [\ion{O}{3}] flux was variable
during the 2008 monitoring period.
\end{abstract}


\begin{keywords}
galaxies: active -- galaxies: nuclei -- galaxies: Seyfert --
techniques: spectroscopic
\end{keywords}



\section{Introduction}

The method of reverberation mapping \citep{blandford1982} uses the
time delay between continuum variations in an active galactic nucleus
(AGN) and the corresponding variations in the flux of broad emission
lines to measure the size and structure of the broad-line region
(BLR). In nearby Seyfert 1 galaxies and low-redshift quasars, the
radius of the BLR measured via reverberation mapping typically ranges
from a few light-days to a few light-months \citep{kaspi2000,
  peterson2004, bentz2009}. Accurate measurement of reverberation lags
requires monitoring campaigns with a cadence sufficient to resolve the
flux variations, and with a total duration at least a few times
greater than the lag time and preferably much longer.

The narrow-line region (NLR) in AGN spans a much larger radial range
than the BLR. Narrow-band imaging and spectroscopic mapping indicate
that the NLR often extends to $\sim$kpc scales in nearby Seyfert
galaxies \citep[e.g.,][]{mulchaey1996,schmitt2003,bennert2006,fischer2013},
although some AGN such as NGC 5548 show evidence of a compact NLR core
in which a significant fraction of the [\ion{O}{3}] emission
originates from scales of just tens of parsecs or less
\citep{kraemer1998}. Recent work has shown that the flux of the
      [\ion{O}{3}] emission line in the well-studied AGN NGC 5548 does
      respond to long-term changes in continuum luminosity: over a
      $\sim20$-year span the [\ion{O}{3}] emission in NGC 5548 faded
      by $\sim20\%$ in response to a decrease in ionizing flux
      \citep{peterson2013}.  Based on this detection of narrow-line
      flux variability, \citet{peterson2013} concluded that the
      majority of the [\ion{O}{3}] emission originates from scales of
      $r\approx1-3$ pc, a significantly smaller size than had
      previously been assumed.  Further measurements of [\ion{O}{3}]
      variations in additional AGN can potentially help to constrain
      the structure of the inner NLR on spatial scales too small to
      resolve directly.

Recently, \citet{zhang2016} claimed to find evidence for [\ion{O}{3}]
variability over a two-month timescale in the Seyfert 1 galaxy Mrk
142, using archival data from the 2008 Lick AGN Monitoring Project
\citep[LAMP2008;][]{bentz2009}. They based this claim on correlations
they found between the fluxes of the [\ion{O}{3}] emission line and
the AGN continuum in LAMP2008 spectra. If true, this result would be
very surprising, as it would suggest that a substantial fraction of
the NLR emission in Mrk 142 is generated on scales smaller than
several light-weeks around the black hole.  The purpose of this paper
is to explain that the claimed detection of [\ion{O}{3}] variability
in Mrk 142 by \citet{zhang2016} is incorrect, as it is based on a
misinterpretation of the data and a misunderstanding of the flux
calibration procedures applied to the data. The LAMP2008 data do not
reveal any evidence for [\ion{O}{3}] variability.

\section{Flux Calibration for Reverberation Mapping Campaigns}

We first review the flux calibration procedures applied to the
LAMP2008 data. The spectra were acquired in Spring 2008 at Lick
Observatory. Very few nights at Lick are photometric, and observing
conditions ranged from thin cirrus to thick cloud cover. A flux
standard star was observed on each night, and the AGN spectra observed
on a given night were calibrated using the same night's standard star
observations. This generally yields a good relative flux calibration,
in which the spectral shape is sufficiently reliable for reverberation
mapping measurements. However, the overall normalization of the flux
scale for each AGN spectrum is essentially random depending on the
relative degree of cloud cover between the standard star observation
and the AGN observations. Even for observations taken in truly
photometric conditions, seeing variations and miscentering of the AGN
in the spectrograph slit will cause some degree of spurious
fluctuations in the broad-line and continuum fluxes measured on
different nights. Thus, some rescaling of the data is required prior
to carrying out measurements of emission-line or continuum light
curves from the spectra.

To normalize the flux scales of the spectra to a consistent scale, we
used the method described by \citet{vgw92}. This method is based on
the assumption that the narrow emission-line fluxes remain constant
over the timescale of an AGN reverberation mapping campaign (typically
a few months), while the continuum and broad emission lines may be
variable. A high-S/N reference spectrum is constructed, and each
individual night's spectrum is scaled in a way that optimally matches
the [\ion{O}{3}] profile of the reference spectrum. The scaling method
employs a linear wavelength shift, a multiplicative flux scaling
factor, and convolution by a Gaussian kernel in order to fit the
nightly [\ion{O}{3}] profiles to the reference spectrum while allowing
for variations in the AGN continuum flux underlying the emission
line. The method does not specifically force each night's spectrum to
have identical [\ion{O}{3}] fluxes after scaling is applied; instead,
it matches the nightly [\ion{O}{3}] profiles to that of the reference
spectrum as closely as possible, which leaves a small level of
residual flux mismatch. \citet{vgw92} stated that ``errors in the
calculated flux scaling factors are generally less than 5\%, and for
most cases much better.''

This residual error can be determined by measuring the fractional
variability amplitude (\fvar) in the [\ion{O}{3}] light curve as
measured from the scaled spectra, because any [\ion{O}{3}] variability
(over and above the amount expected from photon-counting
uncertainties) can be attributed to calibration errors if the line
flux is intrinsically constant.  For a light curve with mean flux
$\langle f \rangle$, variance $\sigma_f^2$, and rms measurement
uncertainty $\delta$, the fractional variability amplitude is
\begin{equation}
\fvar = \frac{\sqrt{\sigma_f^2 - \delta^2}}{\langle f \rangle}.
\end{equation}
Thus, $\fvar^2$ is equivalent to the normalized excess variance in the
light curve.  As an example, for objects observed in the 2011 Lick AGN
Monitoring Project, the values of \fvar\ measured from the
[\ion{O}{3}] light curves after spectral scaling was applied ranged
from 0.5\% to 3.3\% \citep{barth2015}, consistent with the statement
by \citet{vgw92} that the flux scaling errors are often much better
than 5\%.

The LAMP2008 spectroscopic data are available in a public data
release. This includes two versions of the spectra: the ``final''
reduced spectra before scaling is applied, and the scaled spectra
after application of the \citet{vgw92} method.

\section{The LAMP2008 Mrk 142 Dataset}

Our spectroscopic \citep{bentz2009} and photometric \citep{walsh2009}
observations of Mrk 142 in Spring 2008 found a low level of intrinsic
variability in this AGN. The spectroscopic monitoring duration was 68
days. Over this time span, the fractional variability amplitude
\fvar\ was just 2.4\% for the $V$-band light curve and 8.6\% for the
broad \hbeta\ line. Values of \fvar\ below $\sim10\%$ correspond to
fairly weak variability and are not often conducive to measurement of
accurate reverberation lags, and \citet{bentz2009} measured a rather
uncertain lag of $\tau_\mathrm{cen} = 2.88_{-1.01}^{+1.00}$ days by
cross-correlating the \hbeta\ light curve against the $V$-band
continuum.  \citet{bentz2013} present a further discussion of the
LAMP2008 Mrk 142 dataset, noting that the $\sim3$-day lag is
inconsistent with the BLR radius-luminosity relationship, suggesting
that this lag value may be unreliable. More recently, \citet{du2014}
observed Mrk 142 during a period of stronger variability, and found
$\tau_\mathrm{cen} = 6.4_{-2.2}^{+0.8}$ days.

\citet{zhang2016} used the archival LAMP2008 spectroscopic data to
examine the relationship between [\ion{O}{3}] and continuum variations
in Mrk 142. From the outset, it is unlikely a priori that [\ion{O}{3}]
flux variations could be found over a two-month span, given the
expectation that the NLR is extended on scales of at least a few
light-years and possibly much larger. Additionally, the low level of
intrinsic continuum variability in Mrk 142 during the 2008 monitoring
period would be unlikely to lead to any detectable variations in
narrow-line fluxes even if the NLR were very compact.

To search for [\ion{O}{3}] variability, \citet{zhang2016} applied a
spectral fitting procedure to decompose each night's spectrum into
several emission-line and continuum components. The decomposition
procedure was applied to both the unscaled and the scaled
spectra. Then, light curves were measured for the AGN continuum, the
broad Balmer lines, and the [\ion{O}{3}] lines from the model
components of both the unscaled and scaled spectra. They then examined
correlations between the fluxes of different spectral components, in
both the unscaled and scaled data.

Measuring correlations between the fluxes of different spectral
components in the unscaled spectra conveys no useful information about
the behavior of the AGN. As previously described, the overall flux
scales of the nightly spectra are random, resulting from differences
in cloud cover when the AGN and standard stars were observed. In fact,
the greater the variations in cloud cover, the stronger the
correlation that will be seen between the fluxes of different spectral
components, since more variation in the nightly flux calibration will
simply spread the (unscaled) fluxes out over a larger dynamic
range. Using the unscaled spectra, \citet{zhang2016} plot correlations
between the fluxes of \hal\ and \hbeta, between \hal\ and the
continuum, between [\ion{O}{3}] and the continuum, and between
[\ion{O}{3}] and \hbeta. In each case they find a strong correlation
and they quote Spearman rank correlation coefficients. They
specifically claim that the correlations in the unscaled spectra
between [\ion{O}{3}] flux and continuum flux, and between [\ion{O}{3}]
and \hbeta, indicate that ``there is reliable short-term [\ion{O}{3}]
variability over about two months.'' This claim has no merit
whatsoever; these correlations are just the trivial result of
night-to-night transparency variations.


\citet{zhang2016} also examine the correlations between fluxes of
different spectral components in the scaled spectra. In each case they
again find correlations, but these correlations are weaker than those
seen in the unscaled spectra. The reason for this can be understood by
considering the situation of an AGN that shows no intrinsic
variability. For a non-variable AGN, the unscaled spectra will exhibit
strong correlations between different spectral components due to
differences in cloud cover from night to night. After applying the
spectral scaling procedure, one might expect that the light curves
would exhibit zero variability. However, due to the small residual
scaling errors described previously, some small level of variations
will be present in light curves measured from the scaled spectra. When
applying the \citet{vgw92} method, any errors in flux scaling will
affect all spectral components equally, since each nightly spectrum is
multiplied by an overall flux scaling factor. Thus, any residual
errors in spectral scaling will naturally produce a detectable but
spurious correlation between the fluxes of different spectral
components.

For an AGN with low (but non-zero) intrinsic variability, these same
considerations apply. When the residual flux scaling errors are
similar to or larger than the true AGN variability amplitude, the
light curves of different spectral components will be correlated,
because the flux variations will be dominated by residual flux scaling
errors.  In the more favorable regime of strong AGN variability, when
the spectral scaling method works well and residual flux scaling
errors are small, then the [\ion{O}{3}] flux would be expected to show
little or no correlation with the variable continuum flux or broad
Balmer-line fluxes.

In comparing the \hal\ vs.\ continuum flux correlation between the
unscaled and scaled spectra of Mrk 142, \citet{zhang2016} find a
stronger correlation to be present in the unscaled spectra. Instead of
interpreting this as the result of clouds affecting the unscaled
spectra and residual errors affecting the scaled spectra, they claim
that ``the spectral scaling method is not preferred, otherwise a
stronger linear correlation could be expected from the scaled
spectra.'' This is incorrect for a low-variability object such as Mrk
142 observed through highly variable clouds, a situation where the
opposite outcome (a stronger correlation measured from the unscaled
data) is expected.  Similarly, \citet{zhang2016} claim that the
correlation between \hal\ and \hbeta\ fluxes should be stronger in the
scaled than in the unscaled spectra ``if the scaling calibration
method were reasonable''. This statement is based on the same
incorrect reasoning. Again, the exact opposite conclusion should be
reached for the case of an AGN with low intrinsic variability observed
through highly variable cloud cover.

\begin{figure}
\scalebox{0.4}{\includegraphics{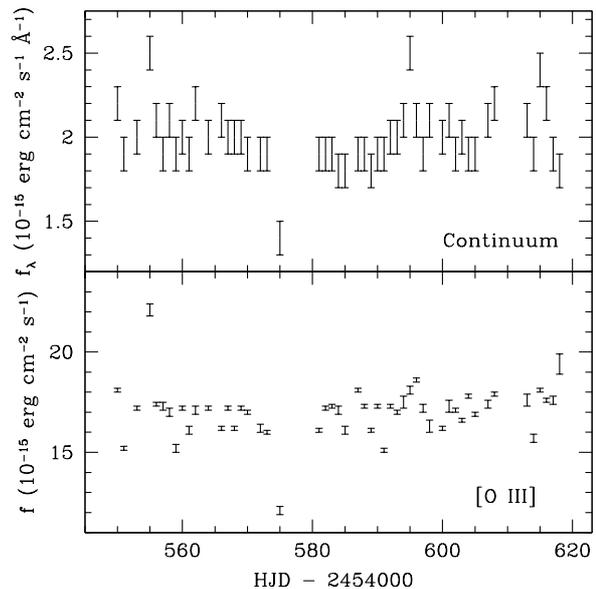}}
\caption{Light curves of the 5100 \AA\ continuum (upper panel) and the
  [\ion{O}{3}] line (lower panel). Data points are from the tables
  presented by \citet{zhang2016}, which are based on their
  measurements from spectral decompositions of the LAMP scaled
  spectra. The apparent discreteness in the light curves is due to the
  fact that the data tables given by \citet{zhang2016} only list
  fluxes to one decimal place of precision (e.g., $2.2\pm0.1$). This
  rounding-off of data points does not affect any of the conclusions
  of this paper.  There are two very noticeable outliers in the
  [\ion{O}{3}] light curve, corresponding to particularly noisy
  spectroscopic observations. The corresponding points in the
  continuum light curve are clearly correlated outliers.}
\label{fig:lightcurves}
\end{figure}

We can easily test whether the correlation between [\ion{O}{3}] and
continuum flux in the scaled spectra, as found by \citet{zhang2016},
is caused by residual flux scaling errors. First, we display in Figure
\ref{fig:lightcurves} the light curves of [\ion{O}{3}] and the
continuum, as listed in Table 1 of \citet{zhang2016}. (Our own
measurements are slightly different, but these small differences do
not affect the outcome of our argument.)  The plot confirms that the
the AGN exhibited very low variability amplitude during this period in
2008. The continuum and [\ion{O}{3}] light curves do not exhibit very
strong features or long-term secular trends, which already disfavors
the possibility that there might be a genuine physical correlation
between the two. The flux variations appear to be dominated by random
scatter, a likely result of errors in spectral scaling.

The scatter in the [\ion{O}{3}] light curve as measured from the
scaled spectra is $\fvar=7.6\%$, significantly worse than the average
for recent reverberation mapping programs. However, a substantial
contribution to this scatter comes from just two obvious outlier
points in the data, at (HJD$-$2454000) = 4555 and 4575. These
measurements correspond to spectroscopic observations having worse
than average S/N, and can be easily recognized as spurious
outliers. If these two points are discarded, \fvar\ is reduced to
5.0\%. Additionally, these same two dates appear as outliers in the
continuum light curve. A scatter of 5\% is still worse than typical,
indicating that this was a somewhat problematic dataset \citep[see
  also][]{bentz2013}.

The presence of strongly correlated errors between the [\ion{O}{3}]
and continuum light curves is obvious from inspection of the light
curves. The problem of correlated errors is well known in
reverberation mapping studies. When emission-line and continuum fluxes
are measured from the same spectra, cross-correlation often yields a
spurious signal at zero lag as a result of the flux scaling errors
affecting both light curves identically \citep{gaskell1987}.  In most
recent reverberation-mapping campaigns (including LAMP2008), the AGN
continuum light curve is measured from photometric data rather than
from the spectroscopic data. This eliminates the problem of correlated
errors between the continuum and emission line light curves, and also
usually produces continuum light curves of higher quality.

\begin{figure}
\scalebox{0.4}{\includegraphics{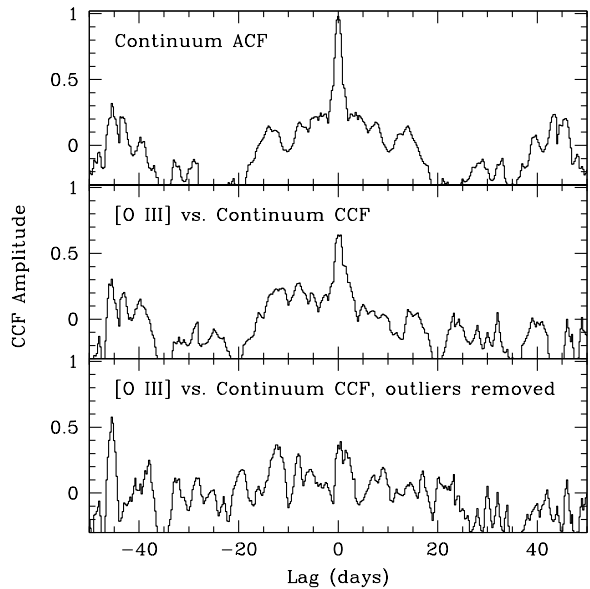}}
\caption{Autocorrelation function of the continuum light curve (top
  panel), cross-correlation of the [\ion{O}{3}] light curve against
  the continuum (middle panel), and cross-correlation after removing
  the two outlier data points from the time series (lower panel). The
  narrow peak in the continuum ACF suggests that the continuum light
  curve is dominated by random errors rather than genuine AGN
  variability. The CCF is also dominated by a narrow spike at zero
  lag, the expected signature for a cross-correlation dominated by
  correlated flux calibration errors between the two light
  curves. After removing the two outlier data points from the light
  curves, there is no significant cross-correlation signal remaining.}
\label{fig:ccf}
\end{figure}

We used the Interpolation Cross-Correlation Function (ICCF) method of
\citet{gaskell1987} to measure the lag between the [\ion{O}{3}] and
continuum light curves as measured from the scaled spectra. The
results are displayed in Figure \ref{fig:ccf}. The autocorrelation
function (ACF) of the continuum light curve exhibits a narrow spike at
zero lag. This is the typical signature of a light curve dominated by
random noise rather than by genuine variability, since real AGN
variability will produce a broader ACF indicating correlations over a
broader range of timescales \citep[e.g.,][]{bentz2009}. Similarly, the
CCF measured between the [\ion{O}{3}] and continuum light curves also
exhibits a sharp peak at zero lag, and no significant signal
corresponding to lags at any longer timescales. This is entirely
consistent with expectations for a CCF dominated by correlated errors
rather than by genuine AGN variability. If we remove the two epochs
corresponding to the strong outliers in the [\ion{O}{3}] light curve, the
resulting CCF has the appearance of random noise, with a weak and
noisy bump near zero lag and no evidence for a genuine lag at any
positive lag time.

If the correlation between [\ion{O}{3}] and continuum fluxes were
real, the CCF would be expected to show evidence for some non-zero lag
between the two light curves, with the lag time indicating the
light-travel time between the continuum emitting region and the inner
NLR.  The observed CCF structure confirms that the correlations found
by \citet{zhang2016} in the scaled spectra are not intrinsic to the
AGN but instead are the result of correlated errors between spectral
components.  There is no evidence for genuine [\ion{O}{3}] variations
in this dataset.

\citet{zhang2016} also find a correlation between the [\ion{O}{3}]
line width and flux in the scaled spectra. This correlation can
similarly be understood as a result of residual errors in the flux
scaling method. The \citet{vgw92} method applies a Gaussian kernel
convolution to each night's spectrum to match the [\ion{O}{3}]
profiles to the reference spectrum, but the profile-matching is never
perfectly realized and there are always residual differences in the
night-to-night profiles. It is not at all surprising that a small
spurious correlation might be introduced into the data by this method,
in which broader widths for the [\ion{O}{3}] profiles after scaling
would correlate with higher fluxes.

\section{Discussion and Conclusions}

As we have shown, the Mrk 142 data from LAMP2008 are somewhat
problematic in that the AGN variability is low and residual scaling
errors are relatively large. This is a dataset that is intrinsically
ill-suited to sensitive investigations of low-level AGN variability,
and it is certainly not suitable for detection of narrow-emission line
variations, which are expected on physical grounds to be extremely
small on timescales as short as two months.

Searching for narrow-line variability in AGN remains an interesting
problem, but the expected variability timescales would typically range
from years to decades, corresponding to the light-travel time across
the core of the NLR. \citet{eracleous1995} estimated the $e$-folding
time for decay of [\ion{O}{3}] emission from a single cloud of density
$n_\mathrm{H}=500$ cm$^{-3}$ to be $\sim4$ years if the ionizing
photon illumination of the cloud were turned off abruptly. Combined
with the spatially extended size of the NLR, is is expected that
short-term fluctuations in ionizing continuum luminosity would be
largely washed away in the integrated response of the NLR. However, a
long-duration secular increase or decrease in ionizing flux could
produce a response in the [\ion{O}{3}] line that might be suitable for
crude reverberation mapping, as demonstrated the recent study of
[\ion{O}{3}] variations in NGC 5548 by \citet{peterson2013}.
Narrow-line flux variations have also been found recently in some
``changing-look'' AGN in which the ionizing continuum luminosity
changes dramatically, such as Mrk 590 \citep{denney2014} and Mrk 1494
\citep{barth2015}.

It is worth considering whether it is even possible in principle to
detect [\ion{O}{3}] variations in a short-duration reverberation
mapping program such as LAMP2008, given that very few nights during
the campaign were photometric. Aside from the [\ion{O}{3}] line, there
is no other reference in the data that can be used to normalize the
flux scales of the nightly spectra; [\ion{O}{3}] is the only strong
narrow line available. The starlight fraction will change from night
to night as a result of seeing variations, and in any case starlight
features are so weak as to be nearly undetectable in the Mrk 142
data. It would only be possible to test for [\ion{O}{3}] variations if
some other external calibration were available, such as a nearby
nonvariable star consistently observed along the same long-slit as the
AGN \citep[e.g.,][]{kaspi2000}.  We conclude that the LAMP2008 data do
not have the capability to demonstrate the presence of [\ion{O}{3}]
flux variations at any variability amplitude that would be physically
plausible.

The flux calibration issues related to the Mrk 142 dataset also
motivate the question of whether improved spectral scaling methods can
be developed that might provide better results than the \citet{vgw92}
scaling method. A new scaling algorithm was recently proposed by
\citet{li2014}, and other new approaches should be
explored. Improvements in spectral scaling methodology may become
particularly relevant for new reverberation-mapping surveys using
multi-fiber instruments to target large samples of AGN at higher
redshift \citep{shen2015,king2015}, because reliable flux calibration
for fiber spectra can be much more difficult than for long-slit
observations.

We have made the LAMP2008 spectroscopic data available in hopes that
it will be of use for a variety of investigations.\footnote{The
  LAMP2008 spectroscopic data are currently available at
  \url{http://www.physics.uci.edu/\~barth/lamp.html}.} We encourage
researchers using these spectra to contact us with any questions
regarding the data reductions or calibrations, so that future
misunderstandings may be avoided.

\section*{Acknowledgements}

Research by AJB is supported by NSF grant AST-1412693. MCB gratefully
acknowledges support through NSF CAREER grant AST-1253702 to Georgia
State University. The 2008 Lick AGN Monitoring Project was supported
by NSF grants AST-0548198 (UC Irvine), AST-0607485 (UC Berkeley),
AST-0642621 (UC Santa Barbara), and AST-0507450 (UC Riverside).




\bibliographystyle{mnras}
\bibliography{ms} 




\bsp	
\label{lastpage}
\end{document}